# A Prioritized Access Point Algorithm for 802.11b Networks in a Lossy Environment

A. N. Omara, Sherine M. Abd El-Kader, Hussein S. Eissa, S. El-Ramly

**Abstract**— In recent years, WLAN technology has been gaining popularity around the world with its sub standard 802.11b receiving major deployments in many indoor and outdoor environments. In this article we investigate the performance of IEEE 802.11b infrastructure networks in the lossless and lossy environments by means of a simulation study. Also, this study shows how the FIFO discipline of the 802.11b MAC affects on the global performance when at least one channel is under the influence of the bursty errors. Furthermore, this paper proposes a channel aware backoff algorithm for the Access Point (AP) to prioritize its transmissions and to accelerate the transmissions in the poor radio channels to enhance the performance of the real time applications. The final results of this simulation study showed that the proposed algorithm is able to enhance the throughput and the delay in lossy environment by an average of 49% and 83% respectively.

**Keywords**—802.11b networks, lossy environment, binary exponential backoff, voice traffic, mean opinion score.

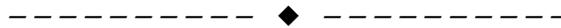

## 1 INTRODUCTION

One of the most critical factors driving the efficiency of the 802.11b networks still remains the ability to overcome the obstacles imposed by the wireless channel. The fairly unpredictable capacity of the links leads to unreliable frame delivery time; hence it dramatically hampers the network performance, especially with regard to real-time applications such as voice.

Beside the performance anomaly of 802.11b networks described in [1] [2] and the channel capture effect [3], the Head-Of-Line (HOL) blocking is very important issue that appears in the lossy environments when the packets addressed to different destinations are queued in FIFO discipline in the MAC layer of the AP. When the HOL packet of MAC suffers from channel errors then the MAC tries to retransmit the packet until it get a response from the recipient or drops the packet if the retransmissions reach to a predefined limit called the retry limit. Those retransmissions of the HOL packet defer the transmissions of the other packet in the same queue regardless of their channel quality. This issue degrades the system throughput whatever the used transport layer protocol (TCP or UDP).

This paper discusses the HOL blocking issue and its effect on the quality of the voice traffic. Also, it proposes a new backoff algorithm that prioritizes the AP in the lossy environment to alleviate this issue.

The rest of the paper is organized as follows. Section 2 defines the 802.11b MAC layer, section 3 discuses the main issues of the 802.11b networks in lossy environment, section 4 presents some efforts performed by researchers that interested in the problem, section 5 discusses the proposed algorithm, section 6 describes in details the implemented scenarios with their obtained results. Finally, section 7 concludes the paper.

This paper discusses the HOL blocking issue and its effect on the quality of the voice traffic. Also, it proposes a new backoff algorithm that prioritizes the AP in the lossy environment to alleviate this issue.

The rest of the paper is organized as follows. Section 2 defines the 802.11b MAC layer, section 3 discuses the main issues of the 802.11b networks in lossy environment, section 4 presents some efforts performed by researchers that interested in the problem, section 5 discusses the proposed algorithm, section 6 describes in details the implemented scenarios with their obtained results. Finally, section 7 concludes the paper.

## 2 PERFORMANCE OF 802.11B MAC

In the IEEE 802.11b [4], the main mechanism to access the medium is the DCF (Distributed Coordination Function), which is a random access scheme based on the CSMA/CA (Carrier Sense Multiple Access with Collision Avoidance) protocol with Binary Exponential Backoff (BEB) algorithm. The DCF describes two access techniques for packet transmission: the basic access mechanism and the RTS/CTS mechanism. In the basic access mechanism, when a WS has a new packet to transmit, it selects a random backoff (in slots) randomly between 0 and CW-1, where CW is the Contention Window. Then, the WS generates a backoff counter, and monitors the channel. The backoff counter of the WS is decremented by 1 if the

————————————————

- *A. N. Omara is with the Department of Computers and Systems, Electronics Research Institute, Egypt.*
- *Sherine M. Abd El-Kader is with the Department of Computers and Systems, Electronics Research Institute, Egypt.*
- *Hussein S. Eissa is with the Department of Computers and Systems, Electronics Research Institute, Egypt.*
- *S. El-Ramly is with the Department of Electronics and Communications, Engineering Faculty, Ain Shams University, Egypt.*

channel is idle for a pre-defined slot time, and frozen otherwise. In the latter case, when the channel is idle for an interval of time that exceeds the DIFS (Distributed Inter-Frame Space) after the channel busy period, the WS resumes the decrement of the backoff counter for every idle slot. When the backoff counter of the WS becomes 0, the packet of the WS is transmitted. After the transmission, if the WS experiences a collision, it generates a random backoff counter for an additional deferral time before the next retransmission. This collision avoidance feature of the protocol intends to minimize collisions during contention among multiple WSs.

The RTS/CTS mechanism involves the transmission of the RTS (Request-To-Send) and CTS (Clear-To-Send) control frames prior to the transmission of the data packet. A successful exchange of RTS and CTS frames attempts to reserve the channel for the time duration needed to transmit the packet under consideration. The rules for the transmission of an RTS frame are the same as those for a packet under the basic access scheme. Hence, the analysis of the basic access mechanism is the same as that of the RTS/CTS mechanism except that the packet transmission times are different in both mechanisms. The RTS/CTS mechanism is more effective than basic access mechanism in case of large packets. It also solves the hidden node problem to a certain extent.

## 3 PERFORMANCE OF FIFO DISCIPLINE IN A LOSSY ENVIRONMENT

The main issues of any wireless technology are related to the nature of the wireless medium. Unlike the wired medium, the wireless medium is open, and all of the signals travelled through the atmosphere are subjected to all sorts of environmental and external factors. This includes storms and the number of walls, ceilings that the signal must pass through.

The lossy channel is that channel whose quality is time varying and may distort the signals travelled through. In 802.11b networks, it is sufficient to have a single lossy channel to produce dramatic performance degradation not only on the lossy channels but also on the lossless channels. In 802.11b networks, the lossy environment raises different issues such as the Performance Anomaly [1] [2], the channel capture effect [3] and the head of line blocking. This paper focuses on the effect of the last issue on the performance of the 802.11b networks.

Many works addressed the HOL blocking issue such as [5]. The author in this paper introduced a simple formula that obtains the expected time ($T_i$) needed to send a packet to $WS_i$. Assuming that ($\sigma$) is the time length of the idle slot in the 802.11b protocol, ($Ts$) is the time required for a successful packet transmission, and ($Tf$) is the time wasted for an unsuccessful transmission, which could be caused by either a packet collision or channel errors. The average value of the $T_i$ could be expressed as shown in equation (1).

$$T_i = (1-p_i)\left[\frac{w-1}{2}\sigma + Ts_i\right] + p_i(1-p_i)\left[\frac{w-1}{2}\sigma + \frac{2w-1}{2}\sigma + Tf_i + Ts_i\right]$$
$$+ ... + p_i^m(1-p_i)\left[\frac{w-1}{2}\sigma + \frac{2w-1}{2}\sigma + ... + \frac{2^m w-1}{2}\sigma + mTf_i + Ts_i\right]$$
$$+ p_i^{m+1}\left[\frac{w-1}{2}\sigma + \frac{2w-1}{2}\sigma + ... + \frac{2^m w-1}{2}\sigma + (m+1)Tf_i + Ts_i\right] \quad (1)$$

Where:
$m$ : is the retry limit.
$w$ : is the minimum contention window.
$p_i$ : is the probability of unsuccessful transmission to the WS $i$.

Note that, ($Ts$) and ($Tf$) are determined by the 802.11b specifications of data rates, PHY header, MAC header, ACK, DCF Inter-Frame Space (DIFS), Short Inter-Frame Space (SIFS), and the length of the payload [6].

Let us consider a system completely managed via the basic access mechanism. Let ($H$) be the packet headers including MAC and physical layer headers, ($P_i$) be the fixed size of all the packets destined to WS$_i$ and ($\partial$) be the propagation delay, then Ts$_i$ and Tf$_i$ could be expressed as shown in equation (2) and (3).

$$Ts_i = H + P_i + SIFS + \partial + ACK + DIFS + \partial \quad (2)$$
$$Tf_i = H + P_i + DIFS + \partial \quad (3)$$

Note that, equation (1) assumes that each WS makes all its retransmissions consecutively; i.e. without any interruption of another transmissions. But this contradicts the fact that all WSs contend the medium and have the same chance to access it; i.e. the WS may wait and froze its backoff counter if there is another transmission from another WS. So, the formula of the $T_i$ is considered the minimum time needed by a WS to send a packet not the maximum time.

Equation (1) represents a simple proof for the HOL blocking when the $T_i$ is the time spent by the AP to send a packet to the destination $i$, then it could be noted that the AP spend more time in transmitting any packet in poor channels (high $p_i$) and this time depends on the number of retransmissions and the contention window length.

## 4 RELATED WORKS

One of the first papers that address the problem of HOL blocking in wireless networks is [7]. It proposes a scheduling algorithm called Channel State Dependent Packet Scheduling (CSDPS). Figure (1) presents the major com-ponents of a CSDPS scheduling system of the base station. A separate queue is maintained for each mobile's packets. Within each queue, packets are served in an FIFO order. Across the queues the service policy could be decided according to the service requirements. In particular, the paper discusses three service policies, namely, Round Robin (RR), Earliest Timestamp First (ETF), and Longest Queue First (LQF).

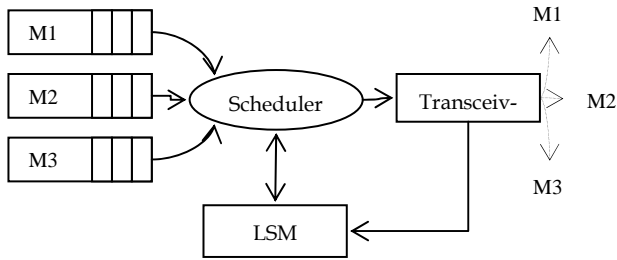

Fig. 1. CSDPS component model

The basic idea of CSDPS is very simple. When a wireless link experiences bursty errors, the scheduling algorithm defers transmission of packets on this link. The Link Status Monitor (LSM) monitors the link states for all mobile hosts. When the LSM determines that a link is in a bad state, it marks the affected queues. The scheduler does not serve the marked queues. The queue is unmarked after a time-out period, which may, for example, be the average link error duration. A link is considered being in a bad state when the acknowledgment for a data packet from the mobile is not received. Simulations show that compared with just using a single FIFO queue for all traffic, CSDPS can achieve much higher data throughput and channel utilization. Since CSDPS alleviates the HOL blocking problem experienced by a single FIFO queue, the average delay of the packets is reduced.

In [5], the author introduced an Adaptive Round Robin (ARR) scheduler which is located at AP's Logical Link Control (LLC) sub-layer. It interacts with the MAC sub-layer to determine which packet (or packet burst) to copy to the MAC. The MAC layer is responsible for transmitting the packet(s) using the 802.11 CSMA/CA [1] multi-access mechanism, except that in this scheme the CW is always fixed at $W$. (i.e., a random number $R$ between $[0, W-1]$, is chosen). For congestion control implementation in ARR, per-station queues, denoted by $Q[i]\ (i = 1, 2, ...n)$, are maintained. Let $P[i]$ denote the HOL packet or packet burst at the front of $Q[i]$. $P[i]$'s are served in a modified round-robin manner. A scheduling penalty is imposed on the queue with a failed transmission to satisfy requirements (a) and (b). Specifically, the ARR scheduler will skip $Q[i]$ for $B[i]$ rounds before the next retransmission. $B[i]$ is increased in a binary exponential manner, depending on the number of retransmission failures. Such a binary exponential backoff is chosen to (1) reduce unproductive transmissions in the possible bursty error period, and (2) emulate the BEB algorithm in the original 802.11 protocol if collisions occur, to prevent network overloading (congestion control). Also, in ARR there is no need to differentiate between collisions and channel errors.

The ARR scheduler guarantees some fairness by using the "Packet Bursty" feature of the 802.11e standard. In this system, a number of packets in a burst could be transmitted successfully before a packet is lost. This will be treated as a "loss event", in which case, the ARR scheduler will immediately postpone the service of the queue to a later round. The successfully transmitted packets in the burst will be removed from LLC layer, and the remaining packets will be transmitted later.

In [8], we introduced an earlier work on the HOL blocking. In this work we introduced two different algorithms for the problem, the Failure Mirror (FM) and the Adaptive Retry Limit (ARL) algorithms. The idea behind the FM algorithm is to reduce the number of packets within the AP MAC which are destined to the bad channels, whereas the ARL algorithm reduces the number of retransmissions in the bad channel. For the FM algorithm, the AP blocks the source's transmissions until the channel of the final destination is available or in good status. This procedure is temporary and active only during the period of burst errors. Unlike the FM algorithm, the ARL algorithm accepts all source's transmissions and reduces the transmissions within the bad channel by minimizing the ceiling of the retransmissions. The final results of [8] proved that, both algorithms enhanced the performance of the network significantly, but the losses increased dramatically.

All of the solutions described above ignored the effect of the BEB on the HOL packet transmission. Also, all of them assumed that the main issue behind the HOL blocking is the number of transmissions addressed to the bad channel. So, the CSDPS scheduler defers the transmissions addressed to the bad channel until its status be good, and the ARR scheduler skips the queue assigned for the bad channel using a round skip scheme similar to the BEB. Furthermore, the previous schedulers suffer from fairness issues, so they need additional schemes for fairness guarantee such as IWFQ [9] and CIF-Q [10]. Like previous schedulers, both solutions, FM and ARL introduced in [8] ignored completely that the effect of the BEB on the performance of the network operates in a lossy environment. Both of the algorithms try to reduce the number of transmissions addressed to the bad channel, either by limiting the retry limit as in ARL, or by blocking the original senders to send more packets as in FM algorithm.

## 5 THE PRIORITIZED ACCESS POINT ALGORITHM

Unlike the algorithms discussed in section 4, the proposed algorithm considers the effect of the BEB on the transmissions of the AP. The Prioritized Access Point (PAP) algorithm assumes that each source in the network sends a stream of unicast packets through the AP. As shown in Figure (2), the AP classifies the sources and the destinations into two groups according to the channel status of their destinations. The Sources group ($S_g$) corresponds to the Destinations group ($D_g$) whose channels are in a good status, and the Sources group ($S_b$) corresponds to the Destinations group ($D_b$) whose channels are in bad status.

The operation of the PAP algorithm is based on two discrete modifications on the MAC of the AP. The first modification accelerates the transmissions of the AP using a new backoff technique called 1.X Exponential Backoff (1.X EB) algorithm instead of the BEB in the 802.11b standard. The second modification which called Blocked DAta (BDA), prioritizes the transmissions of the AP and the good channels status sources over the bad channels status sources.

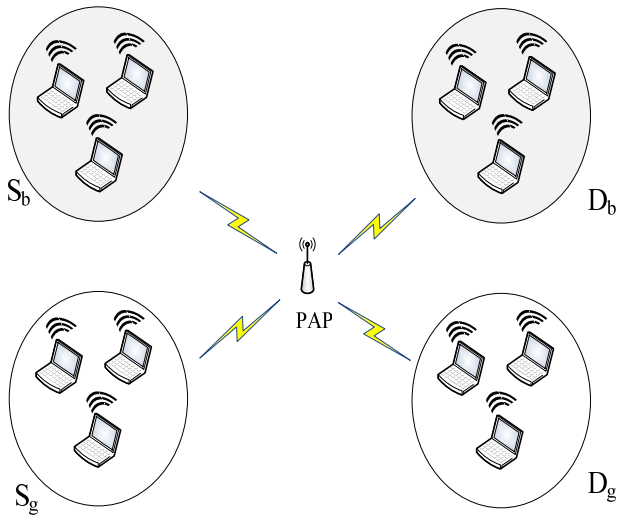

Fig. 2. The centralized network of the prioritized access point

### 5.1 1.X Exponential Backoff Algorithm

The 1.X EB algorithm increases the contention window size smoothly with respect to the BEB algorithm. In BEB algorithm, the AP selects a random slot uniformly between 0 and $CW - 1$, where $CW$ is the contention window size that depends on the status of the last transmission and the order of the retransmission [4]. The BEB algorithm increases the contention window size by multiplying the last size by 2 if the last transmission failed due to the channel errors, and this increment continue until the contention window reaches its maximum size which is 1023 slots as defined in the standard [1], as presented in equation (4).

$$CW_n|_{BEB} = 2\,CW_{n-1} + 1, \quad \text{for } n \geq 1, CW_0 = 31 \text{ slots} \quad (4)$$

Where:
$CW_n$ : is the contention window size of current transmission.
$n$ : is the number of retransmission.

The BEB algorithm is analyzed in [6] analytically by using two dimensional markov chain model, and the obtained results have proved that the chance of a WS to access the medium is inversely proportional to the current contention window size. So, the 1.X EB algorithm will modify the contention window size to vary smoothly compared with the BEB. The 1.X EB algorithm, replace the multiplying factor of equation (4) by another factor Z that is less than 2 and more than or equals 1, as shown in equation (5).

$$CW_n|_{1.X\,EB} = Z \times CW_{n-1} + 1, \quad \text{for} \quad 1 \geq Z \geq 2 \quad (5)$$

When the next transmission is addressed to a bad channel, the value of factor Z should be reduced at each iteration by 0.1 until it reaches its minimum value which is equals to 1 then it will become constant. On contrary, if the next transmission is addressed to a good channel Z value should be increased by 0.1 until it reaches its maximum value which is equals to 2 then it will become constant.

To illustrate the operation of the 1.X EB algorithm an example has to be taken, suppose an example of four successive transmissions in a bad channel followed by two successive transmissions in a good channel followed by six transmissions in a bad channel, assume that the CW is initialized at $CW_0$ at the first transmission of any packet, and the value of Z starts from 2. For the first three transmissions in the bad channel, the CW will have three different sizes as shown below:

$CW_0 = 31 \text{ slots} \qquad\qquad\qquad , Z = 2$
$CW_1 = (2 - 0.1) \times 31 + 1 \approx 60 \text{ slots} \quad , Z = 1.9$
$CW_2 = (1.9 - 0.1) \times 60 + 1 \approx 109 \text{ slots} \quad , Z = 1.8$
$CW_3 = (1.8 - 0.1) \times 109 + 1 \approx 186 \text{ slots} \quad , Z = 1.7$

At this moment, and after four transmissions in the bad channel the value of Z equals to 1.7. For the next two transmissions in the good channel, each transmission is considered a transmission of a new packet so the CW is constant at $CW_0$ and the value of Z increases until it reaches 1.9. For the next six transmissions in the bad channel, the value of Z starts from (1.9 – 0.1) and the

$CW$ will have six different sizes as shown below:

$CW_0 = 31 \text{ slots}$ , $Z = 1.8$
$CW_1 = (1.8 - 0.1) \times 31 + 1 \approx 54 \text{ slots}$ , $Z = 1.7$
$CW_2 = (1.7 - 0.1) \times 54 + 1 \approx 87 \text{ slots}$ , $Z = 1.6$
$CW_3 = (1.6 - 0.1) \times 87 + 1 \approx 132 \text{ slots}$ , $Z = 1.5$
$CW_4 = (1.5 - 0.1) \times 132 + 1 \approx 186 \text{ slots}$ , $Z = 1.4$
$CW_5 = (1.4 - 0.1) \times 186 + 1 \approx 243 \text{ slots}$ , $Z = 1.3$

A question may come to mind, why the value of Z isn't fixed and equals to 1 to minimize the CW size for ever and give the AP more chance to access the medium? The answer to this question was in [14], the author proved analytically that the collision probability is inversely proportional to the size of the CW. So, this paper recommends utilizing a higher contention window after a successful transmission instead of resetting it to $CW_0$. But, this recommendation may be used by the terminals not by the AP, because the higher contention window reduces the transmission probability that may raise the HOL blocking problem within the AP. So, the 1.X EB algorithm proves that the contention window must increase after the failed transmissions to reduce the collision probability and the algorithm also clarifies that this increment must vary smoothly to keep the chance of the AP to access the medium.

### 5.2 Blocked Data Algorithm

The main idea of BDA algorithm is to prioritize the transmissions of the access point and the good channels status sources over the bad channels status sources. So, to achieve this technique the AP should acknowledged all of the frames sent from the $S_g$ sources without exception, and blocked any data frame (the AP will not acknowledge them) sent by the $S_b$ sources temporarily. It should be noted that the AP will block the transmissions of any member belongs to the $S_b$ group until the AP sends a new frame to any destination or the number of successive blocked frames for the $S_b$ member is less than the retry limit by one. The maximum number of successive blocked frames must be less than the retry limit by on to prevent the source from dropping the frame. Those imposed losses at the $S_b$ sources triggers the BEB at their MAC layers (as discussed in section 5.1). Also, those successive retransmissions reduce the transmission probability of them and grant the chance of medium access to the AP and the $S_g$ sources according to the analysis of the BEB discussed in [6] and [14].

It should be noted that the BDA algorithm differs from the FM algorithm described in [8]. In the FM algorithm, the AP doesn't consider the number of successive blocked data frames that may increase the dropping rate at the $S_b$ sources when the successive blocked frames is greater than the retry limit of the MAC. While the BDA algorithm, adapts the successive blocked frames (as described in section 5.2) such that they don't increase the retry limit of MAC.

## 6 SIMULATION STUDY

The simulations are done by using the OPNET 14.0 modeler network simulator. All of the simulation scenarios concentrate on the performance of the 802.11b MAC in lossless and lossy environments. Also, they concerned with the QoS of the voice traffic in the same environments. So, this study selects a common voice encoder to be applied which is G.726 with encoding rate of 32kbps. Note that, the speech activity is selected to be 80% and the Framing Interval (FI) will be variable in our simulation studies. Also, the voice sessions will be one way, i.e. the voice session has one source and one destination.

For the topology of the simulated 802.11b network, it has been chosen to be Basic Service Set (BSS) that uses a central coordinator which is an AP. Also, this simulation study is concerned with the basic DCF mode only, which uses DATA-ACK handshaking only.

In addition to the foregoing, both short and long retransmission limits are initially assigned to be equals to 7 and 4 respectively as they are the default of the standard. On the other side, due to the header effects of fragmentation on the whole performance of the BSS network, the value of Fragmentation Threshold is set to (NONE). Also, all the MAC buffers are assumed to have infinite buffers to avoid any losses due to the limited buffers.

In respect of the geographical size of the network, it has been set to 100 x 100 meters. Therefore, all senders and receivers are within the transmission range of each other and not more than one transmission can take place at a time, and the total bandwidth is divided between the competing transmissions. Note that, the transmission range of the 802.11b protocol is approximately 100 m.

For the physical layer, all fixed nodes are equipped with an IEEE 802.11b PHY operating at frequency 2.4 GHz with a transmission power of 1mW and the physical data rate was selected to be 11 Mbps with High-Rate Direct Sequence Spread Spectrum technology (HR-DSSS).

Finally, the metrics that have been studied are:
- *The throughput*: which refer to the total number of bits received successfully at the MAC of the destination in a period of time.
- *The delay*: it measures the elapsed time from the packet that is queued in the buffer of the source until it is successfully received at the destination. So, it will include the queuing delay and the propagation delay. Note that the voice quality is acceptable for end to end delay less than 150 ms and for some ap-plications the voice quality is acceptable for delay less than 400 ms [11].
- *The packet loss*: it measures the total number of discarded packets in a period of time. Note that the voice quality is acceptable for loss ratio below 5% [12].
- *The Mean Opinion Score (MOS)*: it provides a numerical measure of the quality of human speech at the destination end of the circuit. The scheme uses subjective tests (opinionated scores) that are mathematically averaged to obtain a quantitative indicator of the system performance. Now, the MOS value can be measured using objective methods. Note that, the MOS value ranges from 1 (worst) to 5 (best) and the quality of voice is acceptable if the MOS values are above 3.6 [13].

### 6.1 Simulation Scenarios

Our simulation's study are divided into two groups of scenarios, the primary goal of the first group is to evaluate the performance of the contention access mechanism of the DCF protocol in lossless environment, and to determine the voice capacity according to the maximum achievable throughput and the acceptable values for the predefined metrics. In this group of scenarios, the voice sessions are varying from 1 to 30 sessions and the FI of G.726 is set to 0.02, 0.04 sec respectively.

The second group is divided into two subsets of scenarios. The first subset, studies the impact of the lossy environment on the performance of the contention access scheme. In this subset, the offered load is set to high load which is equal to 75% from the maximum load and to a medium load which is equal to 50% from the maximum load respectively. Note that, the expression of "maximum load" means the maximum load that can be offered without affecting the voice quality. In these scenarios, the voice session is considered failed session if the destination channel suffers from a Frame Loss Ratio (FLR) of 50% or 70% and the FLR is assumed to be constant during the simulation run. While the second subset of scenarios, studies the impact of the PAP algorithm on the performance of the DCF access scheme. All of the simulations are run for 600 seconds and each scenario is repeated for ten times to precise the average values of the scenario.

### 6.2 Simulation Scenarios

This section discusses and analyzes the final results of the simulation scenarios described in section (6.1).

#### 6.2.1 Performance of 802.11b in Lossless Environment

Figure (3), shows the average throughput achieved under varying one way G.726 sessions that transmitting symmetric voice frames. For both values of FI, it could be seen that the throughput increases linearly with the number of voice sessions until it reaches its maximum value then the network is overloaded and the throughput reduces somewhat linearly. For FI equals to 0.02 sec, the throughput is gradually increased from 39 kbps to 550 kbps then it decreases gradually, but for FI equals to 0.04 sec, it is increased gradually from 33 kbps to 864 kbps. From the shape of the graph, it could be noticed that the network is saturated for the number of voice sessions greater than 14 sessions with FI equals to 0.02 sec, while in case of FI equals to 0.04 sec, the network is saturated for number of voice sessions greater than 26 sessions.

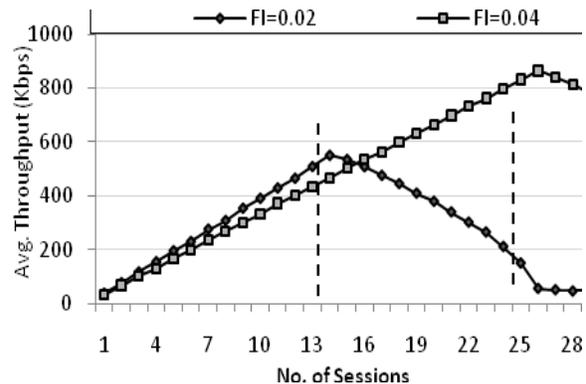

Fig. 3. Avg. throughput in lossless environment

Figure (4) demonstrates the average end-to-end delay of 802.11b in lossless environment. For FI equals to 0.02 sec, from the figure it should be notice that the average end to end delay increases somewhat linearly by on average 28 ms per voice session till the network is saturated then it increases rapidly, while for FI equals to 0.04 sec, the average end to end delay increases linearly by on average 36 ms per voice session till the network is saturated then it increases rapidly also. From the figure also it could be seen that, the average delay is less than the acceptable level (i.e., 150 ms) as long as the number of sessions is less than 13 and 25 sessions for FI equals to 0.02 and 0.04 sec respectively.

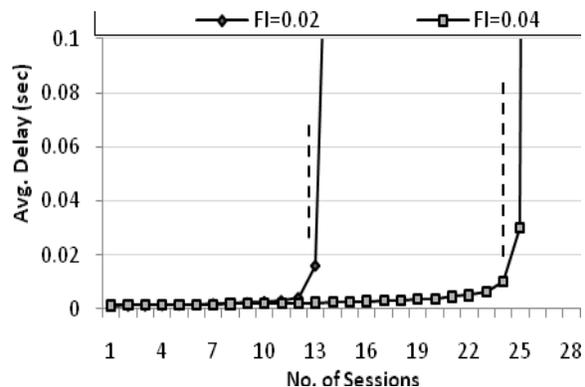

Fig. 4. Avg. end-to-end delay in lossless environment

For average packet loss, as illustrated in figure (5), no packet loss in the lossless environment till the number of voice sessions exceed the level of saturation. It should be noted that, although the packet loss is equal zero the collisions are found and increase until the successive retransmissions exceed the retry limit of the MAC then the losses appear.

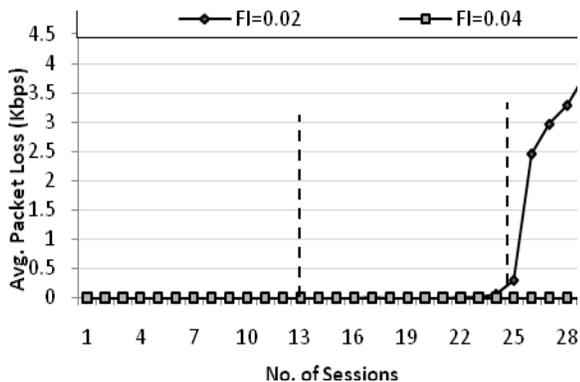

Fig. 5. Avg. packet loss in lossless environment

Figure (6) illustrates the simulated average MOS values. For FI equals to 0.02 sec, the MOS values are always above 3.6 till the number of sessions reach 13 sessions. While for FI equals to 0.04 sec, the MOS values are above the same level till the number of sessions reach 24 sessions. For MOS values greater than 3.6, the quality of voice call is considered acceptable, so the value of 3.6 is very important to determine the maximum load or the voice capacity of the 802.11b network. So, from the figure it could be concluded that for FI equals to 0.02 sec, the voice capacity is equal to 13 voice sessions, while for FI equals to 0.04 sec, the voice capacity is equal to 24 one way voice sessions.

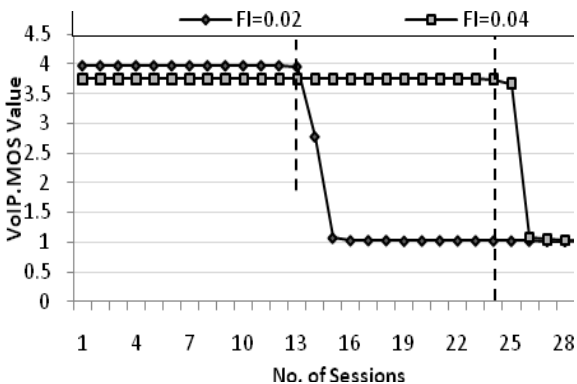

Fig. 6. Avg. MOS values in lossless environment

### 6.2.2 Performance of 802.11b in Lossy Environment

This section demonstrates the simulation results for the lossy environment before and after applying the PAP algorithm. Initially, the scenario is applied for two different offered loads which are 75% and 50% from the maximum load that refer to high and medium load respectively.

From the first scenarios set, for FI equals to 0.02 sec, the high load refers to 10 voice sessions, and the medium load refers to 7 voice sessions. While for FI equals to 0.04 sec, the high load refers to 18 voice sessions, and the medium load refers to 13 voice sessions.

#### 6.2.2.1 High Load Scenarios

Figures (7), (8), (9) and (10) shows the simulated average throughput, delay, losses and MOS respectively for the high loads with FI equals to 0.02 sec.

Figure (7) shows the average throughput of the 802.11b in lossy environment. The average throughput is approximately equal to 387 kbps until the number of bad sessions equal to 3 sessions with 50% FLR then it degrades somewhat linearly by on average 27 kbps per failed session. While for 70% FLR, the average throughput is approximately equal to 387 kbps until the number of bad sessions exceed 1 session then it degrades linearly by on average of 34 kbps per failed session. For the PAP algorithm, it is clear that the average throughput of the PAP algorithm is improved significantly by on average of 29%, 96% for 50% and 70% FLR respectively. Also, it could be seen that the throughput reaches its maximum value when the number of bad sessions reach 7 sessions with 50% FLR, and 3 sessions with 70% FLR.

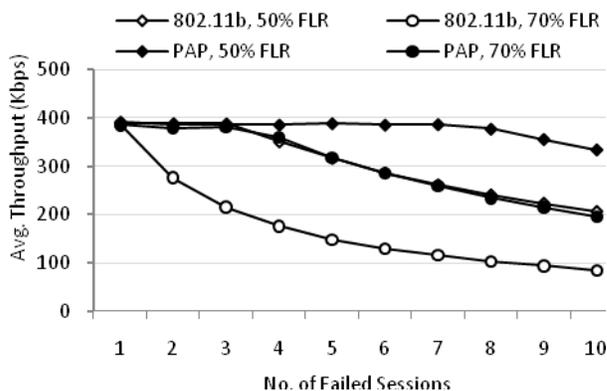

Fig. 7. Avg. throughput (High Load, FI = 0.02 sec)

The average end to end delay results are illustrated in figure (8). Form the figure, it is clear that the 802.11b delay is below the acceptable level as long as the number of failed sessions is less than or equal 3 sessions with 50% FLR. While for 70% FLR, it was sufficient for 1 bad session to increase the delay dramatically and exceed the acceptable level. For the PAP algorithm, it could be seen that the average delay is improved significantly by on average 88%, 73% for 50% and 70% FLR respectively. Also, for 50% FLR, it is clear that the delay is below the acceptable level in case of the number of failed sessions is less than or equal 7 sessions. While for 70% FLR, the delay is below the acceptable level in case of the number of failed sessions is less than or equal 3 sessions.

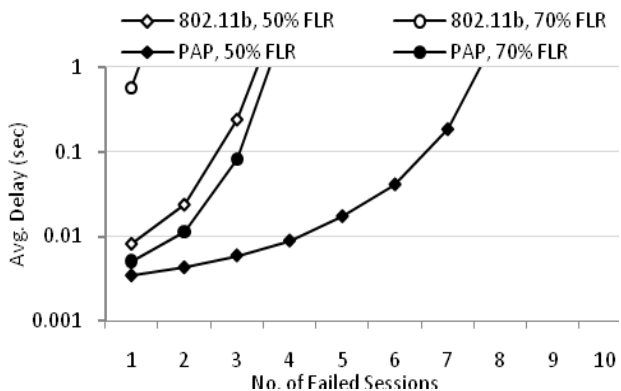

Fig. 8. Avg. end-to-end delay (High Load, FI = 0.02 sec)

Figure (9) shows the packet loss results for both of the 802.11b and the PAP algorithms. For the 802.11b results, it could be noticed that the loss increases rapidly for few number of bad sessions then it increases smoothly for large number of bad sessions; i.e. the increasing rate is not linearly and the amount of packet loss per session decreases when the number of bad sessions increase. This phenomenon appears due to the HOL blocking that blocks all accumulated packets for long time and postpone their transmissions which increase the delay and reduce the losses. If we assumed that the increasing rate is linear, it could be seen that the losses increase by on average 0.2 kbps per failed session with 50% FLR, while for 70% FLR, the losses increase by on average 0.7 kbps. While in PAP algorithm, it could be seen that the losses increased dramatically by on average 48%, 136% for 50% and 70% FLR respectively. Also, it could be seen that the PAP algorithm increase the average losses per failed session to be 0.3 kbps for 50% FLR, and 2 kbps for 70% FLR.

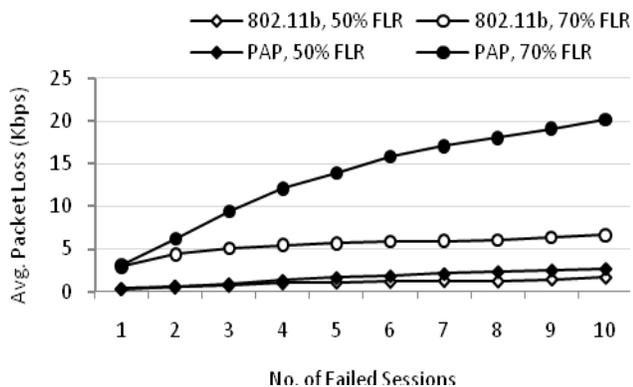

Fig. 9. Avg. packet loss (High Load, FI = 0.02 sec)

Figure (10) illustrates the MOS values for both the 802.11b and the PAP algorithms. For the 802.11b, it is clear that, the MOS values are above 3.6 till the number of bad sessions reach 2 sessions with 50% FLR. While for 70% FLR, it could be seen that one failed session is sufficient to reduce the MOS value to 2.4. Whereas the MOS values in PAP algorithm are improved and always above 3.6 until the number of bad sessions reach 6 sessions with 50% FLR. While for 70% FLR, it is above the same level till the number of bad sessions reach 3 sessions.

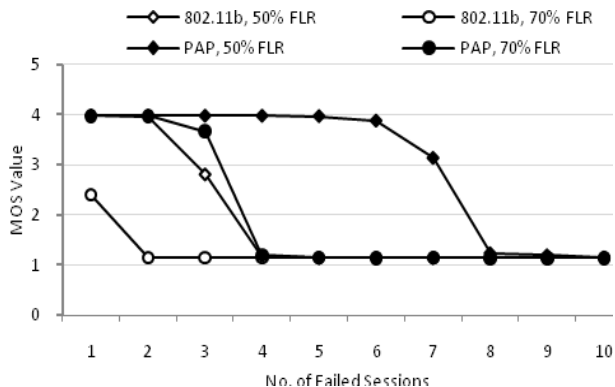

Fig.10. Avg. MOS values (High Load, FI = 0.02 sec)

Figures (11), (12), (13) and (14) show the simulated average throughput, delay, packet loss and MOS respectively for the high loads with FI equals to 0.04 sec.

As shown in figure (11), for the 802.11b results, the throughput is approximately 600 kbps until the number of bad sessions reach 6 sessions with 50% FLR then it degrades somewhat linearly by on average 22 kbps per failed session. While for 70% FLR, the throughput is approximately 600 kbps until the number of bad sessions are more than 2 sessions then it degrades somewhat linearly by on average 28 kbps per failed session. In case of the PAP algorithm, it is clear that the throughput is improved significantly by on average 24% and 83% for 50% and 70% FLR respectively. Also, it is obvious that the throughput is increased until the number of bad sessions reach 14 sessions with 50% FLR, and 6 sessions with 70% FLR.

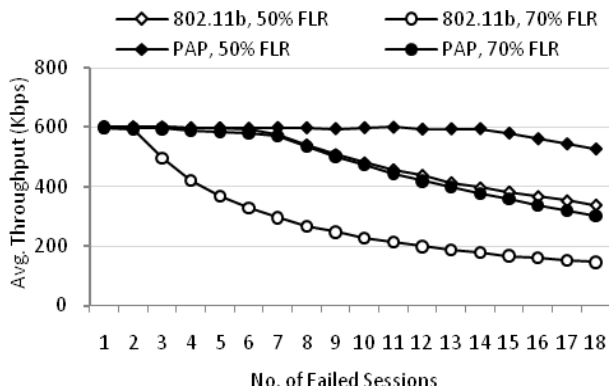

Fig. 11. Avg. throughput (High Load, FI = 0.04 sec)

Figure (12) illustrates the end to end delay results. For the 802.11b, it is clear that the delay is below the acceptable level as the number of failed sessions is less than or equal 6 sessions with 50% FLR. While for 70% FLR, the delay is below the acceptable level as long as the number of bad sessions don't exceed one session. Whereas the average delay is improved significantly by on average 85% and 73% for 50% and 70% FLR respectively in case of the PAP algorithm. Also, for 50% FLR, it is obvious that the delay is below the acceptable level as the number of failed sessions is less than or equal 13 sessions. While for 70% FLR, the delay is below the acceptable level as long as the number of failed sessions is less than or equal 6 sessions.

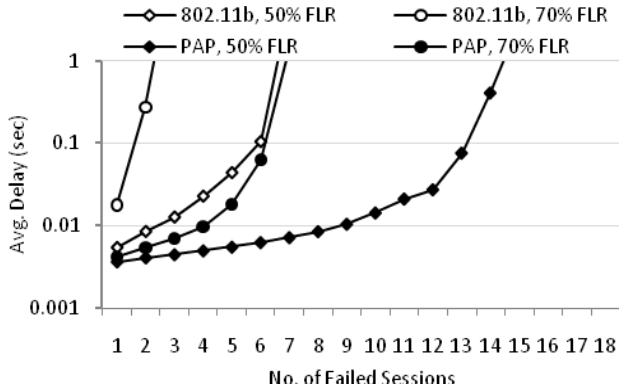

Fig. 12. Avg. end-to-end delay (High Load, FI = 0.04 sec)

The packet loss results for the 802.11b and PAP algorithms are presented in figure (13). For the 802.11b results, it should be noted that the packet loss increases by on average 0.12 kbps per failed session with 50% FLR, while for 70% FLR, the losses increase by on average 0.54 kbps. In case of the PAP algorithm, the losses increased dramatically by on average 70% and 148% for 50% and 70% FLR respectively. Also, it could be seen that the PAP algorithm increases the average losses per failed session to be 0.3 kbps for 50% FLR, and 2 kbps for 70% FLR.

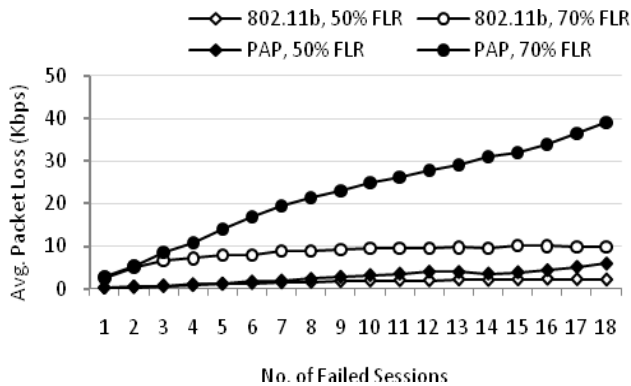

Fig. 13. Avg. packet loss (High Load, FI = 0.04 sec)

Figure (14) illustrates the MOS values for the 802.11b and PAP algorithms. For the 802.11b, the MOS values are above 3.6 till the number of bad sessions reach 5 sessions with 50% FLR. While for 70% FLR, only one failed session keeps the MOS values above the 3.6 level. In PAP algorithm, the MOS values are improved and always above 3.6 till the number of bad sessions reach 12 sessions with 50% FLR. While for 70% FLR, it is above the same level till the number of bad sessions reach 5 sessions.

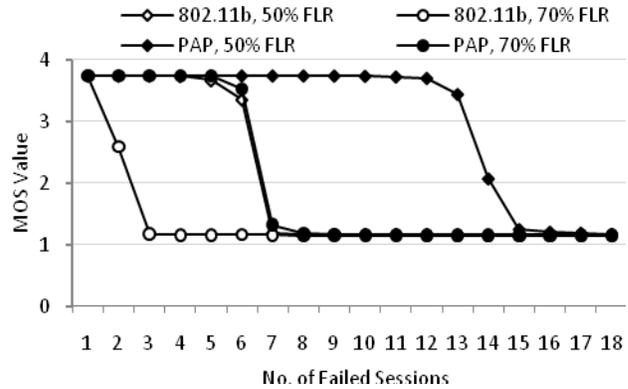

Fig. 14. Avg. MOS values (High Load, FI = 0.04 sec)

### 6.2.2.2 Medium Load Scenarios

Figures (15), (16), (17) and (18) shows the simulated average throughput, delay, losses and MOS respectively for the medium loads with FI equals to 0.02 sec.

As seen in figure (15), for the 802.11b results, the throughput is approximately 268 kbps until the number of bad sessions reach 5 sessions with 50% FLR then it degrades somewhat linearly by on average 19 kbps per failed session. While for 70% FLR, the throughput is approximately 268 kbps until the number of bad sessions exceed 1 session then it degrades somewhat linearly by on average 31 kbps per failed session. For the PAP algorithm, it is clear that the throughput is improved significantly by on average 4% and 84% for 50% and 70% FLR respectively. Also, it could be seen that the throughput is maximized till the number of bad sessions reach 7 sessions with 50% FLR, and 5 sessions with 70% FLR.

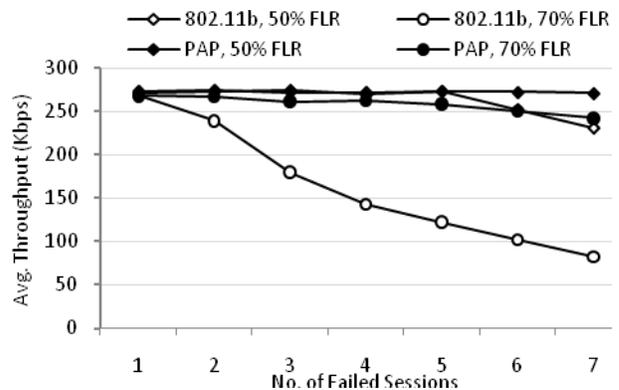

Fig. 15. Avg. throughput (Medium Load, FI = 0.02 sec)

Figure (16) illustrates the average end to end delay results. For the 802.11b, it is clear that the delay is below the acceptable level as long as the number of failed sessions is less than or equal 4 sessions with 50% FLR. While for 70%

FLR, the delay is below the acceptable level as long as the number of bad sessions don't exceed one session. For the PAP algorithm, it could be seen that the delay is improved significantly by on average 82% and 97% for 50% and 70% FLR respectively. Also, for 50% FLR, it is clear that the delay is below the acceptable level as long as the number of failed sessions is less than or equal 7 sessions. While for 70% FLR, the delay is below the acceptable level (150 ms) as long as the number of failed sessions is less than or equal 5 sessions.

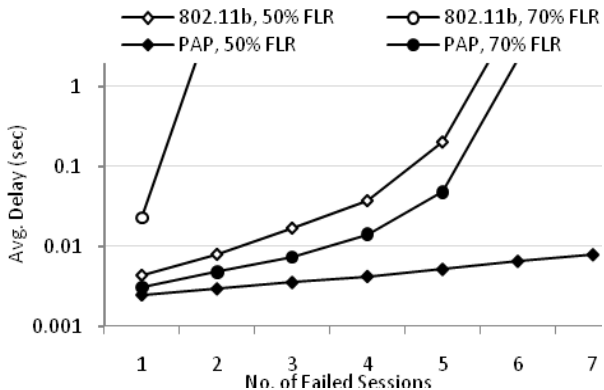

Fig. 16. Avg. end-to-end delay (Medium Load, FI = 0.02 sec)

Figure (17) shows the packet loss results for the 802.11b and PAP algorithm. For the 802.11b results, it could be noticed that the losses increase by on average 0.2 kbps per failed session with 50% FLR, while for 70% FLR, the losses increase by on average 1 kbps. Whereas in PAP algorithm, it could be seen that the losses increased dramatically by on average 15% and 94% for 50% and 70% FLR respectively. Also, it could be seen that the PAP algorithm increases the average losses per failed session to be 0.3 kbps for 50% FLR, and 3 kbps for 70% FLR.

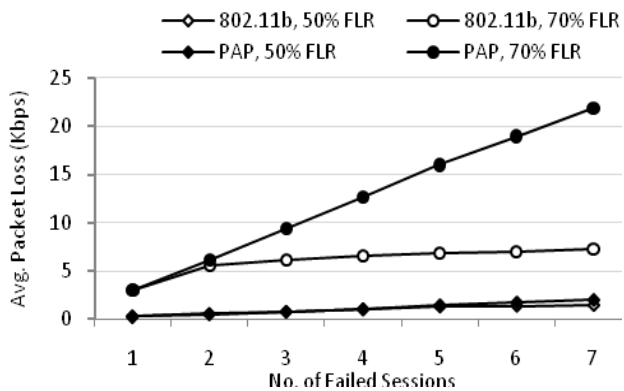

Fig. 17. Avg. packet loss (Medium Load, FI = 0.02 sec)

Figure (18) illustrates the MOS values for the 802.11b and PAP algorithms. For the 802.11b, it is clear that the MOS values are above 3.6 till the number of bad sessions reach 4 sessions with 50% FLR. While for 70% FLR, it could be noticed that only one failed session keeps the MOS above the 3.6 level. While for the PAP algorithm, it could be seen that the MOS values are improved and always above 3.6 till the number of bad sessions reach 7 sessions with 50% FLR. While for 70% FLR, it is above the same level till the number of bad sessions reach 5 sessions.

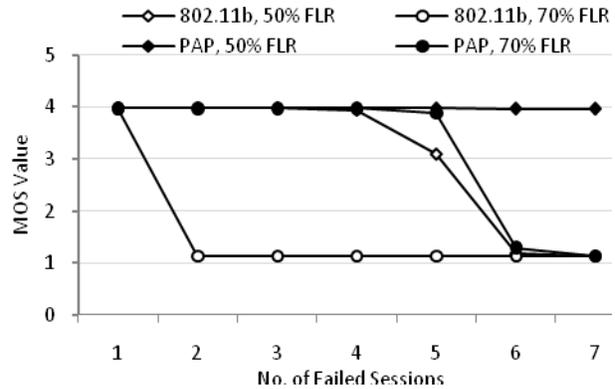

Fig. 18. Avg. MOS values (Medium Load, FI = 0.02 sec)

Figures (19), (20), (21) and (22) shows the simulated average throughput, delay, losses and MOS respectively for the medium loads with FI equals to 0.04 sec.

As shown in figure (19), for the 802.11b results, the throughput is approximately 433 kbps until the number of bad sessions reach 10 sessions with 50% FLR then it degrades somewhat linearly by on average 23 kbps per failed session. While for 70% FLR, the throughput is approximately 268 kbps until the number of bad sessions exceeds 3 sessions then it degrades somewhat linearly by on average 28 kbps per failed session. For the PAP algorithm, it is clear that the throughput is improved significantly by on average 3% and 66% for 50% and 70% FLR respectively. Also, could be noticed that the throughput is increased till the number of bad sessions reach 13 sessions with 50% FLR, and 10 sessions with 70% FLR.

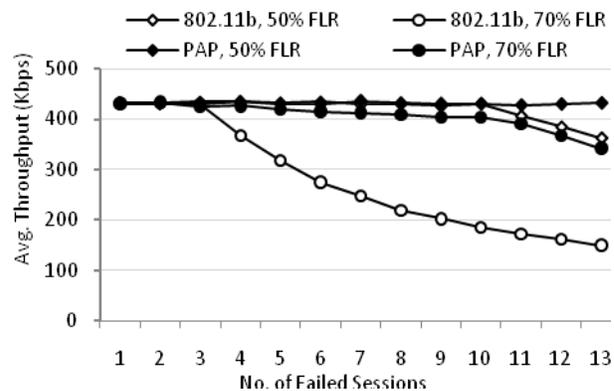

Fig. 19. Avg. throughput (Medium Load, FI = 0.04 sec)

Figure (20) illustrates the average end to end delay results. For the 802.11b, it is clear that the delay is below the acceptable level (150 ms) in case of the number of failed sessions is less than or equal 9 sessions with 50% FLR. While for 70% FLR, the delay is below the acceptable level in case of the number of bad sessions don't exceed 3 sessions. As for the PAP algorithm, it could be seen that the delay is improved significantly by on average 76% and 93% for 50% and 70% FLR respectively. Also, for 50% FLR, it is clear that the delay is below the acceptable level as long as the number of failed sessions is less than or equal 13 sessions. While for 70% FLR, the delay is below the acceptable level as long as the number of failed sessions is less than or equal 10 sessions.

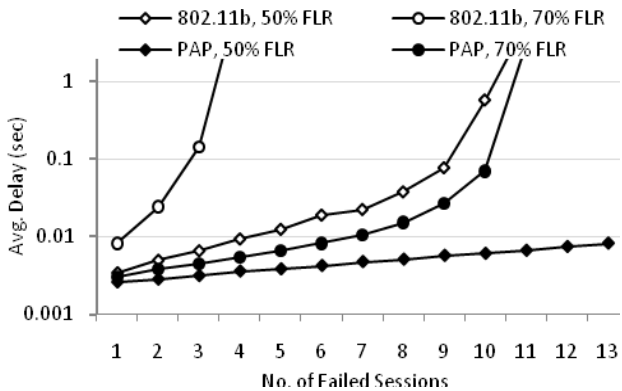

Fig. 20. Avg. end-to-end delay (Medium Load, Fl = 0.04 sec)

Figure (21) shows the packet loss results for the 802.11b and PAP algorithms. For the 802.11b results, it is obvious that the losses increase by on average 0.2 kbps per failed session with 50% FLR, while for 70% FLR, the losses increase by on average 0.9 kbps. While in case of PAP algorithm, it could be seen that the losses increased dramatically by on average 15% and 94% for 50% and 70% FLR respectively. Also, it is clear that the PAP algorithm increases the average losses per failed session to be 0.3 kbps for 50% FLR, and 2.7 kbps for 70% FLR.

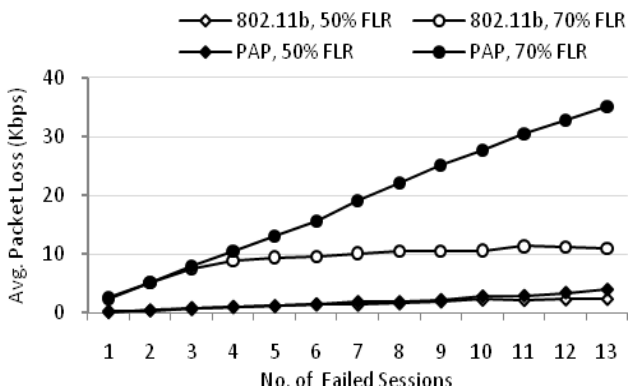

Fig. 21. Avg. packet loss (Medium Load, Fl = 0.04 sec)

Figure (22) illustrates the MOS values for the 802.11b and PAP algorithms. For the 802.11b, it is clear that, the MOS values are above 3.6 till the number of bad sessions reach 8 sessions with 50% FLR. While for 70% FLR, it is clear that only 3 failed sessions keeps the MOS values above the 3.6 level. For PAP algorithm, it is clear that the MOS values are improved and always above 3.6 till the number of bad sessions reach 13 sessions with 50% FLR. While for 70% FLR, they are above the same level till the number of bad sessions reach 9 sessions.

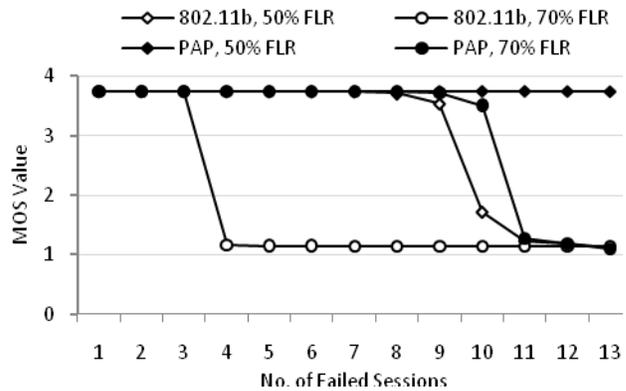

Fig. 22. Avg. MOS (Medium Load, Fl = 0.04 sec)

## 7 CONCLUSIONS

This article proposed a new technique for solving the HOL blocking problem in the 802.11b networks. The main idea behind this solution is to prioritize the transmissions of the AP by increasing the contention window size smoothly after each retransmission in any bad channel.

One of the important conclusions of this work is that, the performance degradation in the lossy environment is similar to that of the overload condition in the error-free environment. But, both performance degradations arise due to different reasons. For overload conditions, there are no free resources for the offered load so all of the terminals are in critical situations and every collision in the network leads the MAC's queues of the flows to be in unbalance state; i.e. their departure rates are less than their arrival rates. But for the lossy environment, the obtained results proved that the performance degrades after certain number of failed sessions not because the lack of resources, but because of the HOL blocking that prioritizes the transmissions of the terminals on the transmissions of the AP which is the bottleneck of the network.

The final results of our simulations study have proved that the PAP algorithm enhances the throughput by on average 58% for the high load, and approximately 39% for the medium load. As for the delay results, the PAP algorithm enhances the average end-to-end delay by on average 80% for the high load, and 87% for the medium load. On the other side, as expected the packet loss increases in the failed sessions only by on average 100% for the high load, but for the medium load it increases by on average 72% and this dramatic increase in the packet loss occurs only due to the effect of accelerating transmissions in the failed sessions. Finally, the MOS values are increased significantly by on average 79% for the high load, and 97% for the medium load, and these increases in the MOS values prove that the AP is capable of serving more failed sessions without affecting the quality of the good sessions.

In the future, it is necessary to enhance the PAP algorithm to guarantee the time and service fairness between the good and failed sessions that could not be achieved in the FIFO discipline. Also, it is necessary to apply this algorithm on different types of traffics to prove its effectiveness.

## BIOGRAPHIES

**Ahmed Naguib Omara** received the B.S. degree in electronics and communications from Shoubra Faculty of Engineering, Zagazig University, in 2003. His research interests include performance evaluation of 802.11 MAC protocol, wireless ad hoc networks, and quality of service. He is a Research Assistant at computers & systems dept., Electronics Research Institute since 2004.

**Sherine Abd El-Kader** has her M.Sc., and Ph.D. degrees from the electronics & communications dept. & computers dept., faculty of engineering, Cairo University, at 1998, and 2003, respectively. Dr. Abd El-kader is an Associate Prof., at computers & systems dept., Electronics Research Institute (ERI). She is supervising three Ph.D students, and 5 M.Sc. Students. Dr. Abd El-kader had published more than 15 papers in computer networking area. She is working in many computer networking hot topics such as; Wi-MAX, Wi-Fi, IP Mobility, Active Queue Management, QoS, Wireless sensors Networks, Ad-Hoc Networking, real-time traffics, Bluetooth, and IPv6. She is an Associate Prof., at Faculty of Engineering, Akhbar El Yom Academy from 2007 till now. Also she is a technical reviewer for many international journals. She is heading the Internet and Networking unit at ERI since 2003. Dr. Abd El-kader is supervising many Graduation Projects from 2006 still now. Finally, Dr. Abd El-kader is the main researcher at two US-EG joint funded projects with University of California at Irvine, CA, USA.

**Hussein Eissa** had his B.Sc. and M.Sc. degrees from electronics & communications dept., faculty of engineering, Cairo University at 1993 and 1996. Dr. Eissa had his Ph.D degree from electronics & communications dept., faculty of engineering, Cairo University in cooperation with electrical engineering dept., university of Pennsylvania, Philadelphia, USA at 2000. Dr. Eissa had an international certificate in business & management from IESES business school, university of Navarra, Spain at 2004. Dr. Eissa is an associate prof. at computers & systems dept., Electronics Research Institute. R. Eissa had published 22 papers in the computer networking area. Dr. Eissa is the director of the R&D sector for the Egyptian IPv6 task force (E-IPv6 TF) since 2004. Dr. Eissa is the director of Information System dept. at ministry of communications & information technology.

**Salwa Hussein Abdelfattah Elramly** is now Professor Emeritus with the Electronics & Communications Engineering Department, Faculty of Engineering, Ain Shams University. She is specialized in Communications Engineering& Signal Processing. Dr. Salwa had her BSc. (1967), MSc. (1972) from the Faculty of Engineering, Ain Shams University, and her PhD from Nancy University, France (1976). She worked as a demonstrator, assistant Professor, Associate Professor, then as a Professor with the Electronics & Communications Engineering Department since1989. She worked as a Director of the Microelectronics Research Unit (attached to Ain Shams Engineering Consultancy Center) for four years, and then was nominated the Head of the Electronics & Communications Engineering Department (2004-2006). Dr. Salwa is an IEEE Senior Member and is the IEEE (Egypt section) Signal Processing Chapter Chair. She is the Secretary of the Permanent Committee for nominating Professors   (Universities Supreme Council), the leader of the Communications Group in the Space Program (with the National Agency for Remote Sensing NARSS) and Secretary General of the Egyptian Society Of Language Engineering ISOLE). Dr Salwa supervised many PhD & MSc thesis and has many publications in Journals and conferences. She is also a member in many committees and share in the steering and technical committees of scientific conferences. Dr. Elramly current research areas are speech coders for mobile communications, smart antennas for location position, multimedia and video conferencing, ciphering, direct delivery of Internet services to users by low earth orbit satellite, multi-user detection in CDMA, GPS for network planning, performance evaluation of CDMA 2000, radars, frequency planning for mobile communications, environmental effects of microwaves emitted from mobile base stations, network management in third generation mobile communications, voice over IP networks, space-time coding, digital video broadcasting for mobile TV, software defined radio, and language engineering.